\documentclass[amsmath,amssymb,natbib,superscriptaddress, twocolumn, prl, aps]{revtex4-1}
\usepackage{tabularx}
\usepackage{graphicx}
\begin{document}

\title{Frequency comparison of ${}^{171}$Yb${}^+$ ion optical clocks at PTB and NPL via GPS PPP}

\author{J.~Leute} 
\address{Physikalisch-Technische Bundesanstalt (PTB), Bundesallee 100, 38116 Braunschweig, Germany}
\author{N.~Huntemann} 
\address{Physikalisch-Technische Bundesanstalt (PTB), Bundesallee 100, 38116 Braunschweig, Germany}
\author{B.~Lipphardt}
\address{Physikalisch-Technische Bundesanstalt (PTB), Bundesallee 100, 38116 Braunschweig, Germany}
\author{Chr.~Tamm}
\address{Physikalisch-Technische Bundesanstalt (PTB), Bundesallee 100, 38116 Braunschweig, Germany}
\author{P.~B.~R.~Nisbet-Jones}
\address{National Physical Laboratory (NPL), Hampton Road, Teddington, TW11 0LW, UK}
\author{S.~A.~King}
\address{National Physical Laboratory (NPL), Hampton Road, Teddington, TW11 0LW, UK}
\author{R.~M.~Godun}
\address{National Physical Laboratory (NPL), Hampton Road, Teddington, TW11 0LW, UK}
\author{J.~M.~Jones} 
\address{National Physical Laboratory (NPL), Hampton Road, Teddington, TW11 0LW, UK}
\author{H.~S.~Margolis}
\address{National Physical Laboratory (NPL), Hampton Road, Teddington, TW11 0LW, UK}
\author{P.~B.~Whibberley}
\address{National Physical Laboratory (NPL), Hampton Road, Teddington, TW11 0LW, UK}
\author{A.~Wallin} 
\address{MIKES Metrology, VTT Technical Research Centre of Finland ltd, Espoo, Finland}
\author{M.~Merimaa} 
\address{MIKES Metrology, VTT Technical Research Centre of Finland ltd, Espoo, Finland}
\author{P.~Gill} 
\address{National Physical Laboratory (NPL), Hampton Road, Teddington, TW11 0LW, UK}
\author{E.~Peik}
\address{Physikalisch-Technische Bundesanstalt (PTB), Bundesallee 100, 38116 Braunschweig, Germany}

\begin{abstract}
We used Precise Point Positioning, a well-established GPS carrier-phase frequency transfer method to perform a direct remote comparison of two optical frequency standards based on single laser-cooled $^{171}$Yb$^+$ ions operated at NPL, UK and PTB, Germany. 
At both institutes an active hydrogen maser serves as a flywheel oscillator; it is connected to a GPS receiver as an external frequency reference and compared simultaneously to a realization of the unperturbed frequency of the ${{}^2S_{1/2}(F=0)-{}^2D_{3/2}(F=2)}$ electric quadrupole transition in ${}^{171}$Yb${}^+$ via an optical femtosecond frequency comb.
To profit from long coherent GPS link measurements we extrapolate over the various data gaps in the optical clock to maser comparisons which introduces maser noise to the frequency comparison but improves the uncertainty from the GPS link. We determined the total statistical uncertainty consisting of the GPS link uncertainty and the extrapolation uncertainties for several extrapolation schemes.
Using the extrapolation scheme with the smallest combined uncertainty, we find a fractional frequency difference $y($PTB$)-y($NPL$)$ of $-1.3(1.2)\times 10^{-15}$ for a total measurement time of 67\,h. 
This result is consistent with an agreement of both optical clocks and with recent absolute frequency measurements against caesium fountain clocks.
\end{abstract}

\maketitle
\section{Introduction}
Novel techniques for remote time and frequency transfer such as optical fiber links are being developed to meet the requirements of ultra-stable optical clock comparisons. 
However, time and frequency dissemination with Global Navigation Satellite Systems (GNSS) is more straightforward to implement and has a much wider availability at the moment, especially over intercontinental distances, which are challenging for optical fiber links.
Therefore, GNSS-based frequency transfer techniques will play an important role for frequency comparisons of optical clocks.

At PTB and NPL optical clocks based on laser-cooled single $^{171}$Yb$^+$ ions are being developed.
The ytterbium ion is of particular interest for the realization of optical frequency standards since it provides two narrow transitions with low systematic uncertainty: an electric quadrupole transition (E2) and an electric octupole transition (E3) that are both recommended by the CIPM (International Committee for Weights and Measures) as secondary representations of the SI second.

GPS Precise Point Positioning (PPP) links between distant clocks reach an instability of $10^{-15}$ at 1 day averaging time \cite{ray, droste}. 
For optical clock comparisons, uncertainties in the $10^{-16}$ range and below are desirable, but the required continuous operation for averaging times of several days is still challenging.
Therefore, we used hydrogen masers as flywheel oscillators between the GPS receivers and the optical clocks and developed a numerical approach to estimate the uncertainties resulting from the discontinuous availability of the optical clock data. 

While others have used GPS PPP links to measure absolute frequencies of optical clocks \cite{ferrari, madej, huang}, this is the first direct remote comparison of two optical clocks via a GPS PPP link.
Beside this work, a direct comparison between optical clocks using the carrier-phase two-way frequency and time transfer technique has been reported in ref.~\cite{hachisu}, which in contrast to the comparison described here requires a dedicated satellite link.

\section{Experimental setup}
\begin{figure}[!h]
\centering
\includegraphics[width=0.9\linewidth]{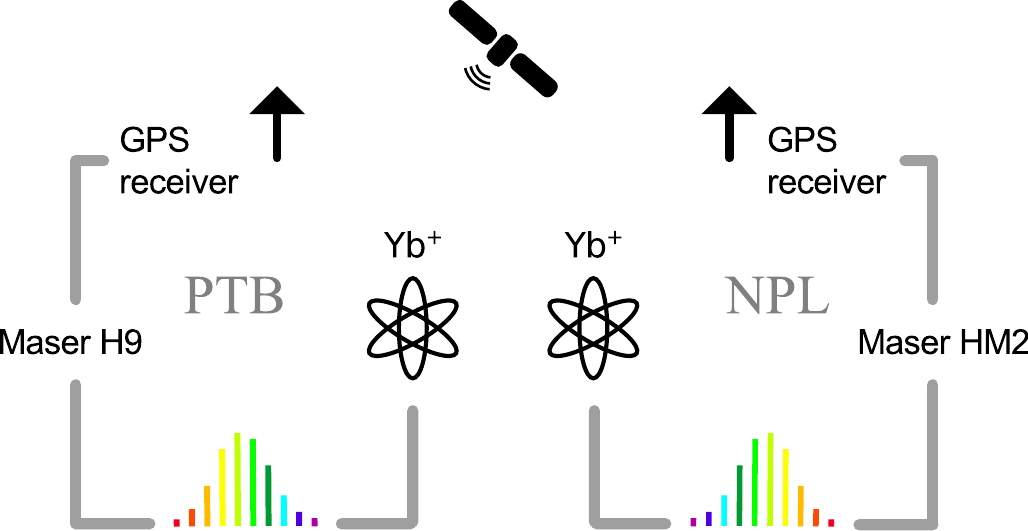}
\caption{Experimental setup at PTB and NPL. A maser serves as an external frequency reference for a GPS receiver and is compared to the optical frequency standard via a frequency comb.}
\label{fig_setup}
\end{figure}
At both institutes an active hydrogen maser serves as a flywheel oscillator; it is connected to a GPS receiver as an external frequency reference and compared simultaneously to the Yb$^+$ frequency standard via a frequency comb (see fig. \ref{fig_setup}). We have in total four measurements: two comparing the masers to the Yb$^+$ standards and two comparing the masers to IGST (timescale of the International GNSS Service).
During continuous operation of all components the maser and IGST contributions drop out when combining all measurements, but this is only true if there are no data gaps or internal dead times as discussed below.  

\subsection{Operation of optical frequency standards}
The frequency of the E2 transition in ${}^{171}$Yb${}^+$ has been realized at both NPL and PTB by stabilizing the interrogation laser to the resonance feature obtained by probing a single trapped ${}^{171}$Yb${}^+$ ion with 30\,ms long Rabi pulses. Details on the experimental setups can be found in refs.~\cite{tamm1, tamm2, godun}.  

At PTB, in contrast to the previously published work \cite{tamm1, tamm2}, the nonlinear frequency drift of the E2 probe laser system has been reduced by about one order of magnitude by prestabilizing its frequency to a more stable reference cavity \cite{huntemann2}. 
Furthermore, since the residual electric field gradients $\nabla E$ are not actively suppressed, the quadrupole shift $\Delta \nu_Q$ that results from the interaction of the quadrupole moment of the excited state with this gradient requires a new evaluation for each measurement. 
The induced frequency shift $\Delta \nu_Q$ depends on the angle that relates the principal axis of the field gradient to the orientation of the quantization axis. 
Due to this relation the shift effect is canceled by averaging the observed transition frequency for three mutually orthogonal orientations of the quantization axis \cite{itano}. 
To determine the magnitude of the quadrupole shift, the magnetic field which defines the quantization axis was switched between two of the three settings every fourth interrogation cycle. In this way, the difference in the quadrupole shift can be measured with quantum projection noise limited uncertainty. Approximately every 8\,h a different combination of orientations was selected. Since the sum over the three directions yields zero, it is straightforward to calculate the individual shifts from the observed frequency differences to be $\Delta \nu_{Q,1}=0.220(6)$\,Hz, $\Delta \nu_{Q,2}=-0.100(6)$\,Hz and $\Delta \nu_{Q,3}=-0.120(6)$\,Hz. The uncertainties result from the statistical uncertainties of the measurements. The systematic uncertainty due to imperfections of the realization of orthogonality is comparable and the frequency shifts are found to be stable within the measurement period. 

At NPL the quadrupole shift $\Delta \nu_Q$ is also canceled by averaging the observed transition frequency over three mutually orthogonal fields. This averaging is performed in a 900\,s cycle with the field settings changed every 300\,s. After each magnetic field step an offset is applied to the clock servo so that it is centered about the respective transition frequency. The servo is allowed to settle for 30\,s before data collection is resumed to prevent any bias from an incorrectly applied offset. The magnitude of the quadrupole shift in each field was obtained by comparing the operational clock with a stable independent system and was found to be 0.7\,Hz, 9.5\,Hz, and -10.2\,Hz. The cancellation of these shifts to the 200\,mHz level is limited by time-variations in the background magnetic field causing uncertainty in the orthogonality of the set of three applied fields. A large (10\,$\mu$T) field is therefore used to minimize the effects of these small fluctuations which causes an increase in the second-order Zeeman shift compared to PTB.

To estimate the instability of the optical frequency standards they were compared to independent stable references. At PTB the probe laser system was measured versus the probe laser system of the strontium lattice clock at PTB \cite{haefner} that shows a relative instability of $\le10^{-16}$ between 1\,s and 1000\,s. The comparison supports an instability of $8\times10^{-15}/\sqrt{\tau(s)}$ for the frequency standard. At NPL the instability was measured by comparing the optical frequency standard against another $^{171}$Yb$^+$ E2 system. An instability of $1\times 10^{-14}/\sqrt{\tau(s)}$ was obtained.

Table \ref{tab:shifts} summarizes the leading frequency shifts and their contributions to the systematic uncertainty for both clocks.
\begin{table}[!t]
\renewcommand{\arraystretch}{1.3}
\caption{Leading relative frequency shifts ${\delta\nu}/{\nu_{0}}$ and the related systematic uncertainty ${u}/{\nu_{0}}$
of the $^{171}$Yb$^+$ E2 transition frequency $\nu_{0}$ during the measurement campaign.}
\label{tab:shifts}
\centering
\begin{tabularx}{\linewidth}{l|XX|XX}
 & PTB & & NPL &\\
\hline
Physical effect & ${\delta\nu}/{\nu_{0}}$ \newline ($10^{-18}$) & ${u}/{\nu_{0}}$ \newline ($10^{-18}$) &${\delta\nu}/{\nu_{0}}$ \newline
($10^{-18}$) & ${u}/{\nu_{0}}$\newline ($10^{-18}$) \\
\hline
Blackbody radiation shift & -524 & 102 & -486 & 99 \\
Second-order Zeeman shift & 968 & 7 & 7566 & 77 \\
Quadrupole shift & 0 & 14 & 0 & 290 \\
Quadratic dc Stark shift & -7 & 4 & -19 & 19 \\
Pathlength error& 0 & 8 & 0 & 6 \\
Servo error & 0 & 3 & 0 & 7 \\
Second-order Doppler shift & -3 & 2 & -5 & 5 \\
Light shift & 0 & 1 & 0 & 8 \\
\hline
Total & 434 & 104 & 7056 & 316 \\
\end{tabularx}
\end{table}

\subsection{Operation of GPS equipment}
A Septentrio PolaRx4TR receiver with NovAtel's GPS-750 antenna and the maser H9 as external time and frequency reference was operated at PTB. A Dicom GTR50 receiver with NovAtel's GPS-702 antenna and the maser HM2 as external reference was operated at NPL. 

\section{Data analysis}
\subsection{GPS PPP}
The comparison of the masers via GPS was done using PPP with the NRCan PPP software \cite{dow} and static IGS orbit and 30\,s clock products \cite{kouba}. In PPP the receiver clock error (receiver clock - IGS time), a tropospheric parameter, carrier-phase ambiguities and the station coordinates are estimated simultaneously and independently for each station. The phase difference between the reference clocks from two stations can be determined by combining their PPP clock solutions so that the reference time scale (IGS time) drops out.

During the comparison period the PPP clock solution of the NPL station shows several steps, corresponding to a cycle slip of a 56\,MHz oscillator on the JAVAD board in the Dicom GTR50 receiver. These steps are corrected for in post-processing of the PPP solution, which is possible because the frequency of 56 MHz is well defined and the time evolution of the clock is unconstrained in the sequential filter used in PPP.
Apart from that the GPS link data is continuous and available from MJD 56951 to MJD 56961.

\subsection{Optical clock data}
By means of the frequency combs, the optical frequency standards were compared to the local hydrogen masers (PTB: H9, NPL: HM2) and then corrected for all systematic frequency shift effects. The measurement of the ytterbium ion clock at PTB started on MJD 56952 at around 20:00 UTC, whereas data from the NPL ytterbium clock is available from MJD 56958 at around 11:00 UTC. The measurements end with the failure of the maser H9 at PTB on MJD 56961 at around 6:00 UTC. At PTB the achieved operational time of more than 130 h was mainly limited by failures of the Yb$^+$ cooling laser and long dark periods of the ion thought to be caused by collisions. At NPL the operational time of 46 h was limited by the occurrence of an abnormally high drift rate of the probe laser's high-finesse stabilization cavity during the first days of the campaign. The uptime of 66\% was limited by the need to repeatedly pause data-taking to measure the background magnetic field.

\subsection{Merging of data sets}
When all components of the experimental setup are operated continuously and there are no data gaps, the merging of the data sets is straightforward. However, both clock data sets comparing the ion clocks to the reference masers contain gaps of various lengths. The data-taking method used at NPL during this campaign leads to a regular dead time of 10\% even if there is no additional failure of equipment or pauses of data-taking for magnetic field background measurements. 

One solution is to only consider time intervals where data from both clocks is available for the frequency comparison. But this results in a fragmentation of the GPS link data, which is unfavorable, since the GPS link instability is higher for shorter averaging times and the fragmentation destroys the phase coherence of the measurement. 
Another solution is to extrapolate the optical clock data to intervals where GPS link data has been recorded for the flywheel oscillators but data from one or both clocks is not available. However, this extrapolation introduces maser noise to the frequency comparison.
The higher GPS link uncertainty due to the fragmentation of the data therefore needs to be balanced against the extrapolation uncertainty caused by maser noise. 

Another challenge for merging the data sets is that the mean fractional frequency differences derived from different fragmentary data sets constitute the mean for different points in time. Therefore maser drifts can affect the frequency comparison. To avoid this we fit a linear polynomial to the fragmentary clock data sets and evaluate the polynomial at the same point in time for all data sets.

\subsection{Uncertainty estimation}
\begin{figure}[ht!]
\centering
\includegraphics[width=1.0\linewidth]{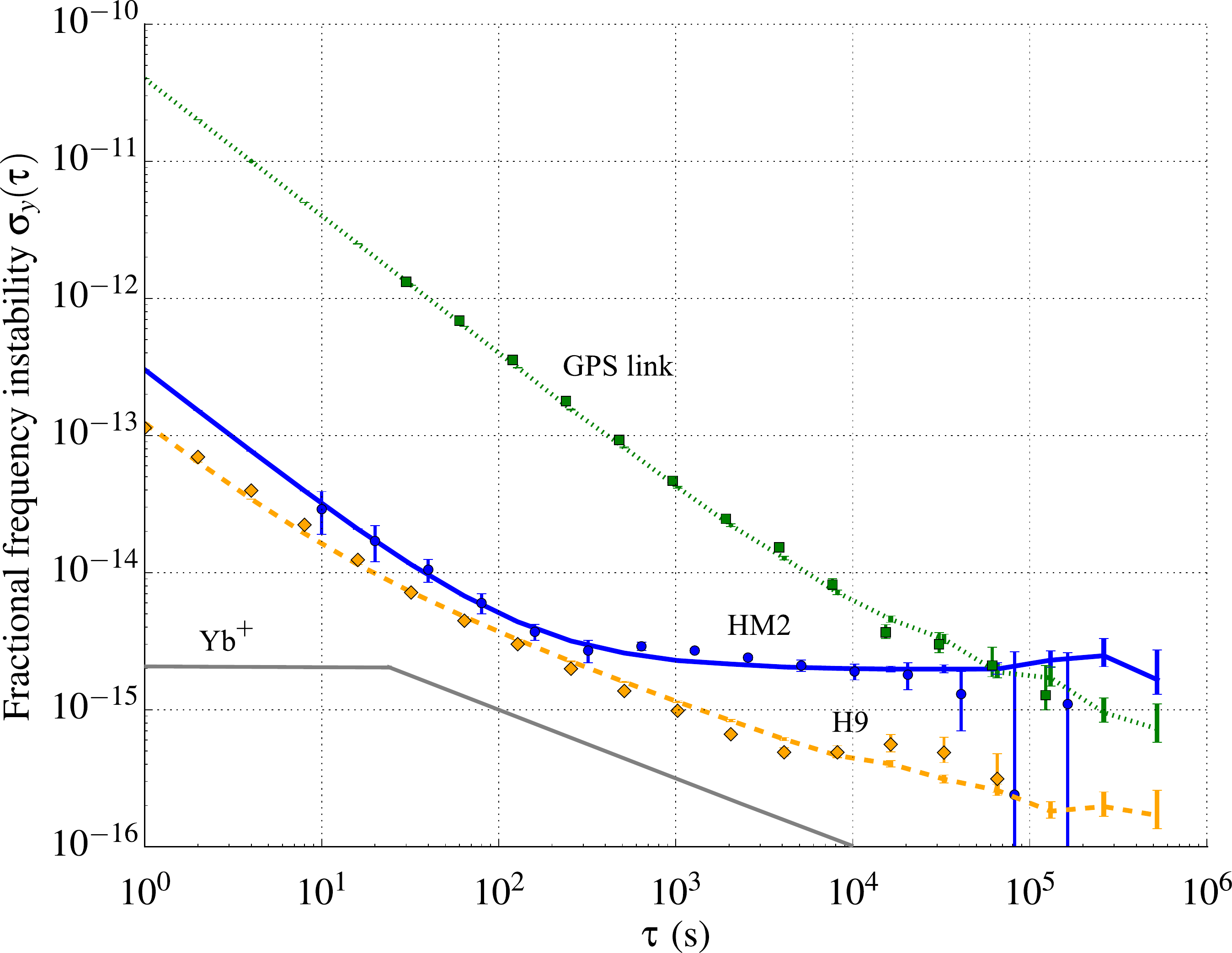}
\caption{Fractional frequency instability from a simulated data set (dotted line: GPS link, solid line: maser HM2, dashed line: maser H9) and experimental data (squares: H9 vs. HM2 via GPS PPP, circles: HM2 three cornered hat analysis vs other NPL masers, diamonds: H9 vs. Yb$^+$ via frequency comb). The grey solid line shows the typical instability of a Yb$^+$ ion clock for comparison.}
\label{fig_simulation}
\end{figure}

We used a numerical approach to estimate the uncertainty associated with the extrapolation of the fragmentary clock data sets and the corresponding uncertainty from the GPS link.
We simulate an ensemble of 1000 data sets with the appropriate noise characteristic of the main contributors, the masers and the GPS link (see tab. \ref{table_simulation}) using a discrete simulation of power law noise \cite{kasdin}. For the determination of the extrapolation uncertainty of the clock data we use the noise characteristic of the maser to which the optical clock is compared, since the optical clock instability is much lower than the maser instability (see fig. \ref{fig_simulation}). Accordingly, for the GPS link uncertainty we use the noise characteristic of the GPS link. In fig. \ref{fig_simulation} the lines indicate fractional frequency instabilities, in each case calculated from a single simulated data set, while the symbols represent experimental data. 

For the uncertainty determination, each of the 1000 simulated data sets is mapped to the fragments of the real data set that it aims to represent and to the extrapolated interval. Then the mean fractional frequency difference of the simulated data set in both mappings is calculated. The uncertainty is estimated by the standard deviation of the difference of the results in both mappings for the ensemble of simulated data sets. 

\begin{table}[ht]
\renewcommand{\arraystretch}{1.3}
\caption{Assumptions about noise characteristics (White Phase Modulation: WPM, White Frequency Modulation: WFM, Flicker Floor: FF) of the masers and the GPS link for the simulation.}
\label{table_simulation}
\centering
\begin{tabularx}{0.8\linewidth}{l|lll}
 & WPM [$s/\tau$] & WFM [$\sqrt{s/\tau}$] & FF \\
\hline
Maser HM2 & $3.0\times10^{-13}$ & $3.5\times10^{-14}$ & $2.0\times10^{-15}$ \\
Maser H9 & $1.2\times10^{-13}$ & $3.5\times10^{-14}$ & $3.0\times10^{-16}$ \\
GPS PPP link & $4.0\times10^{-11}$ & $5.0\times10^{-13}$ & $3.0\times10^{-16}$ \\
\end{tabularx}
\end{table}

\section{Results}
We investigate several extrapolation cases, where all or only some of the gaps are extrapolated. We estimate the extrapolation uncertainty and the GPS link uncertainty for all cases to find the one that leads to the smallest combined uncertainty.
As can be expected, the extrapolation uncertainty depends on the distribution and length of the data gaps and on the instability of the maser. 

\begin{figure}[!h]
\centering
\includegraphics[width=0.85\linewidth]{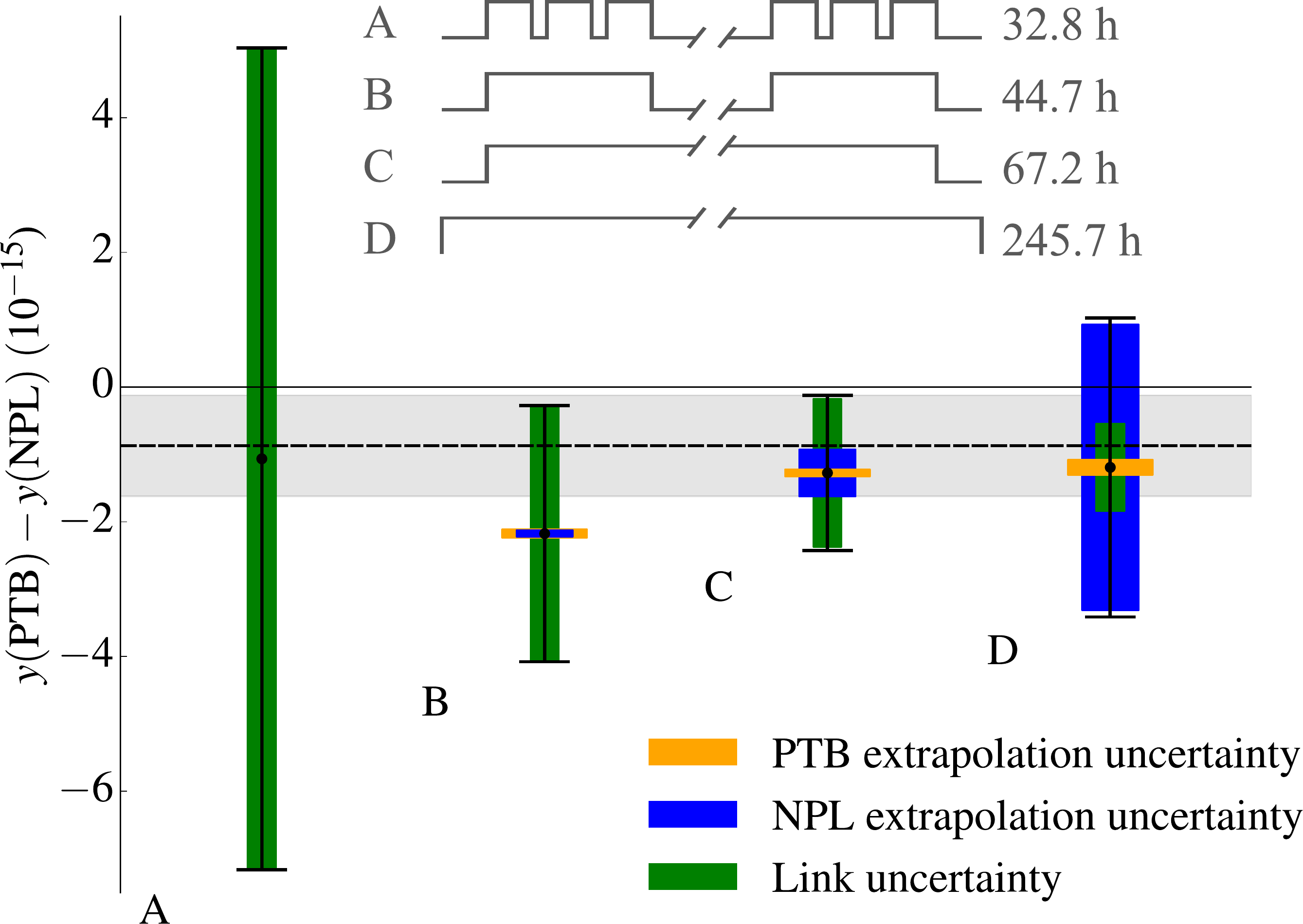}
\caption{Fractional frequency difference $y($PTB$)-y($NPL$)$ of the $^{171}$Yb$^+$ E2 transition with statistical uncertainties for four different extrapolation cases ($A$-$D$). The dashed line and the shaded region show the result from absolute frequency measurements against caesium fountain clocks and its uncertainty \cite{tamm1, godun}. Also shown are sketches of the data structure for the four extrapolation cases and the corresponding total averaging times.}
\label{fig_results}
\end{figure}

In the first case ($A$) no gaps are extrapolated and fragmentation is at its maximum. 
Resulting from the quadrupole cancellation scheme at NPL, dead times of 30\,s are removed from the GPS PPP data every 300\,s. This limits the phase-continuous averaging time and leads to a large statistical uncertainty of $6\times10^{-15}$ for a total averaging time of 32.8\,h (see fig. \ref{fig_results}).

In the second case ($B$) we extrapolate the intrinsic 10\% dead time (30\,s data collection pauses) of the NPL measurement intervals and gaps in the optical clock data at PTB where NPL data was available. As a consequence, small extrapolation uncertainties (significantly below $10^{-16}$) emerge while the GPS link uncertainty is noticeably reduced to $2\times10^{-15}$. This reduction can not be explained solely by the increase in total averaging time from 32.8\,h to 44.7\,h (compare fig. \ref{fig_simulation}) between cases $A$ and $B$, but mainly by the increased duration of the phase-continuous intervals of the GPS data in case $B$. 
 
In the third case ($C$) we extrapolate over the interval starting with the NPL clock data and ending with the failure of the maser H9 at PTB (67.2\,h), thus removing additional gaps during that period from maintenance of the clock systems that occur in both the NPL and PTB optical clock data sets. Here, the NPL extrapolation uncertainty becomes noticeably larger ($3\times10^{-16}$) due to the performance of the NPL maser HM2, while the PTB extrapolation uncertainty is still below $10^{-16}$ and the total statistical uncertainty is ${1.2\times10^{-15}}$.

Finally, we extrapolate over the entire period where GPS data is available (245.7\,h), including several days at the beginning of that period where no NPL Yb$^+$ clock data is available ($D$).
The NPL extrapolation uncertainty dominates the combined statistical uncertainty with $2\times10^{-15}$ resulting from the flicker floor of the NPL maser. 

Besides the statistical uncertainties for the four cases, a systematic uncertainty from the GPS PPP link can be neglected \cite{droste}.

For case $C$ with the smallest combined statistical uncertainty we find a fractional frequency difference
$y($PTB$)-y($NPL$)=-1.3(1.2)\times10^{-15}$, including the systematic uncertainty from the optical clocks (see tab. \ref{tab:shifts}) and a correction for the gravitational redshift corresponding to the height difference of 66.7(4)\,m.
The result is in good agreement with recent absolute frequency measurements against caesium fountain clocks (${-0.87(75)\times10^{-15}}$) \cite{tamm1, godun}.

\section{Conclusion}
The first direct remote comparison of two optical clocks via GPS PPP has been performed.
Using a numerical approach it is possible to estimate the uncertainty from the extrapolation of fragmentary clock data to profit from long coherent GPS link measurements. This enables frequency comparisons with an uncertainty of approx. $10^{-15}$ even with highly fragmentary clock data.
With better flywheel oscillators and a longer measurement period smaller uncertainties are achievable even with a data taking scheme with frequent interruptions. 
Additionally, improved GPS time and frequency transfer techniques like Integer Precise Point Positioning \cite{petit} and more sophisticated weighting schemes for the frequency averages \cite{benkler} will allow further improved frequency comparisons of optical clocks.

\section*{Acknowledgment}
We would like to thank S. H\"afner for providing the probe laser of the strontium frequency standard at PTB.
This work was funded by the EMRP project SIB04 Ion Clock. The EMRP is jointly funded by the EMRP participating countries within EURAMET and the European Union.

\bibstyle{apsrev4-1}
\bibliography{GPS_Yb_comparison}

\end{document}